\documentclass[useAMS, usenatbib, usegraphicx]{mn2e}
\usepackage{amsmath, amssymb, amsfonts, subfigure} 
\usepackage{appendix}

\usepackage[bookmarks, colorlinks=true, citecolor = red]{hyperref}
\usepackage{graphicx}

\newcommand{\beg}{\begin{equation}}
\newcommand{\e}{\end{equation}}
\newcommand{\begs}{\[}
\newcommand{\es}{\]}
\newcommand{\erf}{\text{erf}}
\newcommand{\erfc}{\text{erfc}}

\begin{document}

\title[The MLP Distribution]{The MLP Distribution: A Modified Lognormal Power-Law Model for the Stellar Initial Mass Function}

\author[Basu et al.]{Shantanu Basu$^{1,2}$\thanks{E-mail: basu@uwo.ca}, M. Gil$^2$\thanks{Current affiliation: School of Information Studies, McGill University, Montreal, Canada.} and Sayantan Auddy$^1$ \\
$^1$ Department of Physics and Astronomy, The University of Western Ontario, London, ON, Canada N6A 3K7.\\
$^2$ Department of Applied Mathematics, The University of Western Ontario, London, ON, Canada N6A 5B7.
}

\maketitle

\begin{abstract} This work explores the mathematical properties of a distribution introduced by \citet{Basu}, and applies it to model the stellar initial mass function (IMF). The distribution arises simply from an initial lognormal distribution, requiring that each object in it subsequently undergoes exponential growth but with an exponential distribution of growth lifetimes. This leads to a modified lognormal with a power-law tail (MLP) distribution, which can in fact be applied to a wide range of fields where distributions are observed to have a lognormal-like body and a power-law tail. We derive important properties of the MLP distribution, like the cumulative distribution, the mean, variance, arbitrary raw moments, and a random number generator. These analytic properties of the distribution can be used to facilitate application to modeling the IMF. We demonstrate how the MLP function provides an excellent fit to the IMF compiled by \citet{Chabrier2} and how this fit can be used to quickly identify quantities like the mean, median, and mode, as well as number and mass fractions in different mass intervals. 
\end{abstract}


\section{Introduction}\label{intro}

The distribution of stellar and substellar masses at birth, the Initial Mass Function (IMF), is a key feature of star formation. It has been studied intensely since first estimated by \citet{Salpeter}, who measured a power-law tail for high masses of the approximate form $dN/d\ln M \propto M^{-1.35}$. Subsequent work has established a shallower slope at masses less than $M_{\odot}$ and a turnover at approximately $0.2\,M_{\odot}$ when the masses are placed in logarithmically spaced bins. These compilations of the IMF have tended to identify a power-law profile at high masses \citep[e.g.,][]{Scalo,Kroupa,Kroupa2}, although earlier work \citep{Miller} did fit the IMF with a lognormal distribution. \citet{Chabrier} (see also \citet{Chabrier2}) has compiled an IMF in the substellar and low mass stellar regime and advocates a lognormal fit for masses up to $M_{\odot}$ and a power-law fit for $M > M_{\odot}$. By appealing to the Central Limit Theorem (CLT), \citet{Chabrier} claims a better rationale for the lognormal fit at low masses than the approach of using broken three-component power-laws \citep{Scalo,Kroupa,Kroupa2}. However, Chabrier's overall fit also requires joining a lognormal with a power law at high masses, so has one joining condition instead of two. An ideal next step is to find a single function with no joining conditions that has a rationale that is at least on par with appeals to the CLT. 

There are many disciplines in which a desired distribution is one that is like a lognormal at low and intermediate values, with a characteristic peak and turnover, but transitions to a power law distribution at high values. Besides astronomy, this need has arisen in fields as diverse as biology \citep{Limpert}, computer science \citep{Mitzenmacher}, ecology \citep{Allen}, and finance \citep{Mandelbrot}, to name several. A review of common statistical resources reveals that very few analytic functions of this type exist, hence the attempts to fit empirical distributions by patching together different functions over different domains. Power-law distributions were first introduced by \citet{Pareto} to explain the distribution of incomes seen in data from many different countries. The distribution of city sizes also shows a power-law character and regularity across many countries. It is referred to as Zipf's Law \citep{Zipf}. Henceforth, we refer to pure power-law distributions as Pareto distributions, in conformity with much of the statistical literature. 

In this paper, we analyze and characterize the properties of a hybrid three-parameter probability density function (pdf) introduced by \citet{Basu}. We feel that it can be used to fruitfully model data sets that exhibit both lognormal-like and power-law behavior. Indeed, the MLP distribution 
illustrates the fact that many generative processes that lead to a lognormal distribution, can with some modification, yield a power-law tail instead. As a result of its origin as a modified lognormal, two parameters of the MLP distribution are identified as $\mu_0$ and $\sigma^2_0$, preserving the notation of the lognormal distribution, while the third parameter is $\alpha$, the power-law index that also characterizes the Pareto distribution. However, the parameters $\mu_0$ and $\sigma^2_0$ no longer represent the mean and variance, respectively, of the logarithm of the random variable, as they do for the lognormal distribution. This three-parameter function is a simpler and more readily usable form of a more general four-parameter pdf derived by Reed in several papers \citep{Reed1, Reed2}. The latter function arises from a stochastic growth law, Geometric Brownian Motion, rather than the pure exponential growth used by \citet{Basu}. An advantage of the pdf we use here is that it introduces only one additional parameter beyond that in the lognormal. As a result, it is a natural first step when fitting data that may look like a modified lognormal, and its relatively compact analytic closed-form expression makes it easy to use with common fitting techniques. 

The model of \citet{Basu} falls into one of two major categories of IMF models in modern day star formation theory, that of an accretion based scenario in which temporal effects are important. The basic idea is that star formation is a killed (or stopped) process, with mass accretion terminated by events such as stellar outflows, ejection from the mass reservoir, or any other reason for emptying the mass reservoir. The distribution of accretion lifetimes then plays a key role in setting the shape of the IMF, while the Jeans mass does not. Other models in this category include those of \citet{Adams}, \citet{Bate}, and \citet{Myers, Myers2}, with the latter developing a model that can fit the observed IMF by tuning several parameters. 
The alternate scenario is one in which the mass distribution is determined by the spatial properties of the gas distribution and by gravitational instability, so that some combination of the turbulent spectrum and the Jeans mass set the IMF, e.g., \citet{Padoan,Hennebelle}, with the latter developing an analytic approach. Although the fitting of analytic functions to the IMF cannot alone settle which models may be most suitable, they can however greatly facilitate analysis in fields like galaxy studies, where the conclusions depend strongly on an adopted IMF model. Analytic functions also allow a simpler analysis of the effect of varying IMF parameters. 

The paper is organized as follows. In \autoref{lognormalIntro} and \autoref{paretoIntro} we introduce the lognormal and Pareto distributions, respectively, and present some of their relevant properties. This is done for completeness of the presentation and to add context when reading the new results about the MLP distribution. In \autoref{MLPdist} we discuss the formulation of \citet{Basu} that leads to the MLP distribution. We then examine some relevant mathematical properties of this distribution in \autoref{MLPMathProperties}, which includes expressions for its cumulative distribution function, mean and variance, arbitrary raw moments, and an approximation to its mode. These expressions, excluding the approximate mode, are shown to reduce to the corresponding lognormal expressions in the appropriate limit. In \autoref{FitSec}, we fit the MLP distribution to the IMF of \citet{Chabrier2} and then use the analytic function to quickly estimate some of the above described IMF properties, as well as a cumulative mass fraction. Some closing remarks are given in \autoref{conclusions}. The derivations of the expressions given in \autoref{MLPMathProperties} are included in \autoref{Derivations}, which also contains relevant properties of the error and complementary error functions and related integrals used in this paper. 

\section{ The Lognormal Distribution}\label{lognormalIntro}
According to the CLT of probability theory \citep{Gut}, if $X_1,...,X_n$ are identically distributed, independent random variables with mean $\mu$ and standard deviation $\sigma$, then $Z =  \left(\sum X_i - n\mu\right)/(\sqrt{n}\sigma)$
converges in distribution to the standard normal variable $N(0,1)$ with zero mean and variance of unity. In addition, the identical distribution assumption for $X_1,...,X_n$ may be dropped, and the result will also follow provided certain conditions (Lindeberg's) are satisfied \citep{Gut}. As pointed out in \citet{Golberg} the CLT is used to partially justify why so many observable phenomena appear to be normally distributed: if the variable of interest is thought to be influenced by the sum of a large number of independent factors, then the CLT can be invoked to explain the apparent normality. However, it is frequently the case that the variables of interest are non-negative, whereas normal variables are not. One way to circumvent this complication is to have a distribution with similar properties but which is always positive. Suppose that $Z$ above was instead assumed to be of the form $Z = X_1\cdot X_2\cdot...\cdot X_n$. Then $W = \ln Z = \sum Y_i$, where $Y_i = \ln X _i$. Note that $Y_1,...,Y_n$ satisfy the conditions of the CLT, so that $W$ converges to a normal random variable, $W \sim N (\mu, \sigma^2)$. Moreover, if $W$ is normal, then $Z = e^{W}$ has a lognormal density \citep{Golberg, Aitchison, Crow},  
\beg \label{lognormal} f_Z(z;\mu, \sigma^2) = \frac{1}{z \ \sqrt{2\pi}\sigma}\exp\left(-\frac{(\ln z - \mu)^2}{2\sigma^2}\right),   \ \ \ z>0 \ .\e
Thus, under the assumptions above, $Z$ converges to a lognormal.  We list some of the relevant properties of the lognormal distribution below:

\begin{enumerate}	
\item Raw Moments: \beg \label{momLog}  E[Z^k] = \exp\left(k\mu + \frac{1}{2}k^2\sigma^2 \right) \ ;\e
\item Mode: \beg z_m = \exp(\mu - \sigma^2)  \ ;\e
\item Variance: \beg \label{varLog} \text{Var}(Z) = \exp\left(2\mu + \sigma^2\right)[\exp(\sigma^2) - 1] \ ; \e
\item Cumulative Distribution Function (cdf) \beg  \label{cumLog} F_Z(z;\mu, \sigma^2) = \frac{1}{2}\left[1 +\erf\left( \frac{\ln z - \mu}{\sqrt{2} \sigma}\right)\right] \ ,\e
\end{enumerate}
	where $\erf(z)$ is the error function (see \autoref{erf} in the appendix). 
	\section{The Pareto Distribution}\label{paretoIntro}
Pareto distributions have been used extensively to model a wide variety of phenomena in the sciences and social sciences, such as the size of forest fires, the intensity of earthquakes, the citations to papers, and the population of cities. For a recent review, see \citep{Clauset}.

A random variable $X$ has a Pareto distribution if its probability density function is 

	\beg \label{paretoDens} f(x) =\frac{\alpha b^\alpha}{x^{\alpha+1}} \, , \ \ x > b > 0 \e 
        \citep{Pareto, Mitzenmacher, VonSeggern},
	where $\alpha > 0$. We list some of its more relevant properties below:

	\begin{enumerate}

	\item Raw Moments: \beg E[X^n] = \frac{\alpha b^n}{\alpha-n} \ , \ \ \alpha > n \ ; \e

	\item Mode: \beg x_m = b \ ; \e 

	\item Variance: \beg \sigma^2 = \frac{\alpha b^2}{(\alpha-1)^2(\alpha-2)} \ , \ \ \alpha > 2 \ ; \e 

	\item Cumulative Distribution Function (cdf)

	\beg F_X(x;\alpha,b) = 1 - \left(\frac{b}{x}\right)^\alpha \ . \e 

	\end{enumerate}

	\section{The MLP distribution}\label{MLPdist}
	The key element in the derivation of the MLP distribution is that even though the initial values of a random variable may have a lognormal density, later time evolution could in fact skew their distribution. In other words, the subsequent time of
growth of the random variable (representing a physical quantity) is itself another random variable. Our pdf can be derived analytically on this basis using a few simplifying assumptions.

Imagine that initial values of a quantity $M_0$ are drawn randomly from
a lognormal distribution with parameters $\mu_0$ and $\sigma_0$.
If the subsequent growth of an object with $M_0 = m_0$ 
is characterized by exponential growth
\begin{equation} \frac{dm}{dt} = \gamma\,m, \end{equation}
with fixed growth rate $\gamma$, then the multiplicative relation between values $m$ of individual objects is preserved.
A lognormal distribution is then maintained with the same $\sigma_0$ but the mean of the logarithmic values shifts to $\mu_0 + \gamma t$
after a fixed time $t$. However, we can treat the time as a random variable and draw it from an exponential pdf 
\begin{equation} \label{exponential}
f(t) = \delta \, e^{-\delta\,t},
\end{equation}
where $\delta$ is a stopping rate. In this case, we can derive the probability
density function of final masses $m$ as 
\beg f(m) = \int_0^\infty \frac{\delta e^{-\delta t}}{\sqrt{2\pi}\sigma_0 m}\exp\left(-\frac{[\ln m - \mu_0 - \gamma t]^2}{2\sigma _0^2}\right)dt \ .\e
Using the identity in \autoref{I11} we find the closed form 
\begin{align} f(m) = &\frac{\alpha}{2}\exp\left(\alpha\mu_0 +\alpha ^2\sigma_0 ^2 / 2\right)m^{-(1+\alpha)} \notag \\ &\times \erfc\left(\frac{1}{\sqrt{2}}\left(\alpha\sigma_0 - \frac{\ln m - \mu _0}{\sigma_0}\right)\right) \ , \ m \in[0,\infty) \ , \label{MLP} \end{align}
where $\alpha = \delta/\gamma$.

\begin{figure}
	\centering
	\subfigure{
	\includegraphics[width = 0.47\textwidth]{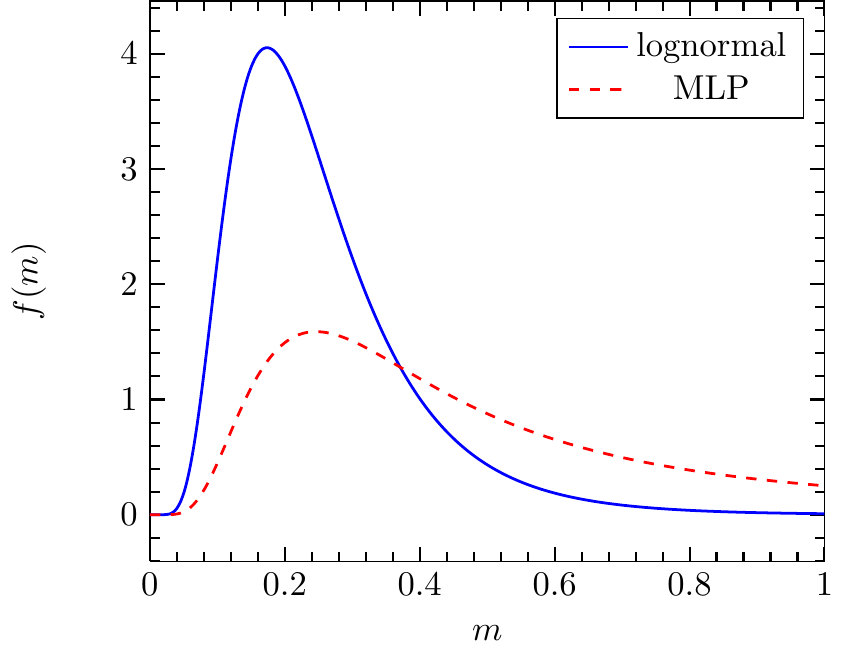}
	}
	\subfigure{
	\includegraphics[width = 0.47\textwidth]{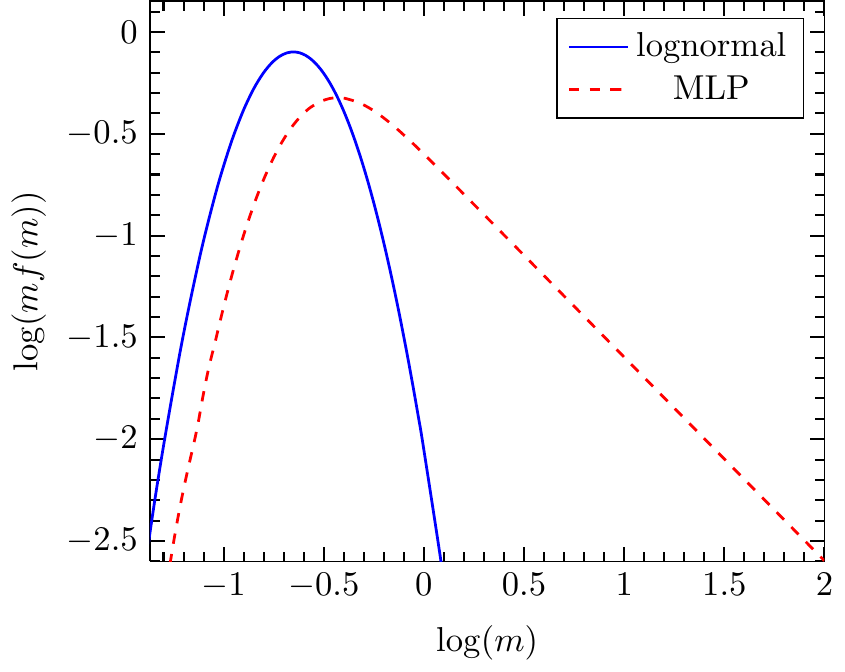}
	}
	\caption{\label{densitiesGenericFig} Comparison of a lognormal density function with $\mu=-1.5$ and $\sigma=0.5$ with the MLP density function with $\mu_0 = -1.5$, $\sigma_0 = 0.5$, and $\alpha = 1$. The mass is normalized by a solar mass $M_{\odot}$.}
	\end{figure}

\autoref{densitiesGenericFig} shows the MLP density $f(m)$ (\autoref{MLP}) and a lognormal distribution for similar values of parameters. The specific values of the parameters used here are somewhat arbitrary, but correspond approximately to a best fit lognormal for the low mass end of a stellar mass distribution, as modeled in \citet{Basu}. While $\mu_0$ is largely a scale-dependent parameter, the other two parameters are expected to fall in the approximate range $0 \leq \sigma_0 \leq 1$ and $1 \leq \alpha \leq 2$ based on fits of lognormal and power-law distributions to a wide range of phenomena in the sciences and social sciences \citep{Limpert,Clauset}. The function derived here is related to a four-parameter distribution \citep{Reed1,Reed2} that can be derived under the assumption of geometric Brownian motion, $dm = \gamma\,m\,dt + a\,m\,dw$, which is a stochastic growth law where $dw$ represents $R(0,1)$, the uniform random variate in the interval $[0,1]$, and $a$ is an amplitude of the fluctuations. The resulting pdf is a double-tailed Pareto distribution, with coefficients that are roots of a quadratic equation. The pdf we derive here results from a simpler model and has the advantage of a single-expression closed form. It may also more easily correspond to many data sets that seemingly warrant only one additional parameter (beyond the original two of the lognormal distribution) in order to get a reasonable fit and quantify the data.
\section{Relevant Mathematical Properties of the MLP Distribution}\label{MLPMathProperties}

A few relevant mathematical properties of the MLP distribution are examined below, leaving most of the necessary calculations in \autoref{Derivations}. Let $M$ denote a random variable with an MLP distribution of density $f(m)$.

\subsection{Cumulative Distribution}
The MLP cumulative distribution function, 
\beg F(m) = \int_{-\infty}^m f(t)dt \ ,\e   
is given by
\begin{align} F(m) &= \frac{1}{2}\erfc\left(-\frac{\ln(m)-\mu_0}{\sqrt{2}\sigma_0}\right) \notag\\
&\hspace{-20pt}-\frac{1}{2}\exp\left(\alpha\mu_0+\frac{\alpha^2\sigma_0^2}{2}\right)m^{-\alpha}\erfc\left(\frac{\alpha\sigma_0}{\sqrt{2}}-\frac{\ln(m)-\mu_0}{\sqrt{2}\sigma_0}\right). \label{cumDist} 
\end{align}
Notice that as $m \rightarrow 0$ we can see by the result in \autoref{lim} that the behaviour of $F(m)$ is dominated by the first term, which is exactly the cumulative distribution function for a lognormal random variable (\autoref{cumLog}) with parameters $\mu_0$ and $\sigma_0$. %\citep{Golberg}. 

\subsection{Mean, Variance, and Raw Moments}
The $k$th raw moment of $M$, defined as the expectation value of $M^k$, 
\beg E[M^k] = \int_0 ^\infty m^k f(m)dm \, , \e
exists if and only if $\alpha > k$, in which case it is given by 
\beg E[M^k] =\frac{\alpha}{\alpha - k} \exp\left(\frac{\sigma_0^2k^2}{2} + \mu_0 k\right) \label{MLPMomemnts} , \ \alpha > k. \e
Note that this expression is exactly the formula for the raw moments of a lognormal distribution (\autoref{momLog}) with parameters $\mu_0$ and $\sigma_0$, scaled by the factor $\alpha/(\alpha - k)$, and in the limit as $\alpha \rightarrow \infty$, the expressions are identical. This is consistent with the derivation of the MLP distribution in \autoref{MLPdist}. The limit $\alpha \rightarrow \infty$ corresponds to $\gamma \rightarrow 0$ for finite death rate $\delta$, so that the drift term vanishes, and the distribution remains a lognormal with mean $\mu = \mu_0$. Assuming $\alpha > k$, we can obtain the following expressions, including the mean and variance of the distribution:
\beg \label{mean} E[M] = \frac{\alpha}{\alpha - 1}\exp\left(\frac{\sigma_0 ^2}{2} + \mu_0\right) , \ \alpha > 1,\e
\beg E[M^2] = \frac{\alpha}{\alpha - 2}\exp\left(2[\sigma_0^2 + \mu_0]\right), \ \alpha > 2,\e
\begin{align} \label{var} \notag \text{Var}(M) &= E[M^2] - (E[M])^2 \\ &= \alpha \exp(\sigma_0^2 + 2\mu_0)\left(\frac{e^{\sigma_0^2}}{\alpha-2} -  \frac{\alpha}{(\alpha - 1)^2}\right), \ \alpha > 2.
\end{align}
Higher moments around the mean can be computed using \autoref{MLPMomemnts} with the identity
\beg E[(M - E[M])^n] =  \underset{j=0}{\overset{n}{\sum}}\binom{n}{j}E[M^{j}](-1)^{n-j}(E[M])^{n-j} \ .\e

\subsection{Mode \label{mode}}

To find the mode, that is, the value $m^*$ that maximizes the MLP pdf in \autoref{MLP}, we must solve the transcendental equation
\beg f'(m) = 0 \iff K\erfc(u) = e^{-u^2} \label{modeEq}  \ , \e
where 
\begin{align} K &= \sigma_0(\alpha + 1)\sqrt{\frac{\pi}{2}}\,, 
&u=\frac{1}{\sqrt{2}}\left(\alpha\sigma_0 - \frac{\ln m - \mu _0}{\sigma_0}\right).
\end{align}
Although the solution to \autoref{modeEq} will generally require the implementation of numerical methods, we note that if $K\approx 1$ then $u = 0$ provides an approximate solution to \autoref{modeEq}, which in terms of the original parameters results in
%, and $u = (B + C\ln m)$.
\beg m^* = \exp\left(\mu_0 + \alpha\sigma_0^2\right) \ . \label{firstApprox} \e
The approximation in \autoref{firstApprox} is useful only when the assumption $K \approx 1$ is closely met, and behaves poorly otherwise. However, when a precise numerical solution is required, one can use this approximation as a starting point in the iteration procedure being implemented. It is worth noting that even if one tried to find the peak in the space $mf(m)$ vs $\log(m)$, the resulting equation to be solved is in fact the same, owing to the fact that $mf(m)$ is just $f(m)$ with a different value of $\alpha$, and $d/d(\log(x)) = x(d/dx)$. 

\subsection{Random Number Generation}
\label{randomgen}

For practical purposes of comparing data sets with a model
distribution, it is valuable to be able to draw a random sample from
the model. Using the definition of the lognormal random variable 
(see \autoref{lognormalIntro}), a random number drawn from a lognormal pdf 
(\autoref{lognormal}) is an element of the lognormal random variate
\beg
\label{lognvar}
L(\mu, \sigma) \sim \exp(\mu + \sigma\,N(0,1) ),
\e
where $N(0,1)$ is the normal random variate with zero mean and variance of unity.
For drawing from an exponential pdf (\autoref{exponential}), the exponential 
random variate is
\beg
\label{exponvar}
E(\delta) \sim - \delta^{-1}\ln (R(0,1)),
\e
where $R(0,1)$ is the uniform random variate in the interval $[0,1]$.
The above formula can be obtained from \autoref{exponential} through the
general method of calculating
the cdf, inverting it, and then drawing the argument as an element of the
uniform random variate $R(0,1)$.
We can use \autoref{lognvar} and \autoref{exponvar} to derive a random variate 
for the MLP distribution. We note that the MLP distribution (\autoref{MLP}) 
is formally obtained from an initial 
lognormal pdf with mean $\mu_0$ under the transformation
$\mu_0 \rightarrow \mu_0 + \gamma\,t$, in which $t$ is chosen randomly
from an exponential distribution. Therefore, we can use \autoref{lognvar}
and \autoref{exponvar} to write that the MLP random variate is
\beg
\label{mlpvar}
M(\mu_0,\sigma_0,\alpha) \sim  \exp(\mu_0 + \sigma_0\,N(0,1) - \alpha^{-1}\ln(R(0,1)) ).
\e

\autoref{mlpvar} can be used to draw a random sample from the MLP pdf
just as \autoref{lognvar} can be used to draw a random sample from a 
lognormal pdf. 

%For small samples, the two distributions may not be distinguishable.
%We have drawn random samples from an MLP distribution using \autoref{mlpvar} 
%and the best fit parameters presented in Section \ref{FitSec}, then
%made histograms with logarithmic bin width of $0.1$. 
%When we fit these histograms with either a lognormal or an MLP distribution, we find that the
%lognormal provides a better fit by the reduced chi-squared criterion whenever the sample
%size is about 100 or less. Since the power-law tail is not well populated unless several hundred 
%elements are drawn from the MLP pdf, and the lognormal pdf has one less fitting parameter, it
%can perform better in goodness-of-fit tests for small sample sizes.

%\autoref{MLP} is conceptually derived from an initial
%lognormal that is then evolved with exponential growth for an 
%exponential distribution of lifetimes. Therefore, we can combine
%the well-known techniques for drawing random samples from lognormal
%and exponential distributions to find that the MLP
%random variate is
%\beg
%M(\mu_0,\sigma_0,\alpha) \sim  \exp(\mu_0 + \sigma_0\,N(0,1) + \gamma E(\delta)),
%\e
%where
%\beg
%E(\delta) \sim -\delta \, \ln (R(0,1))
%\e
%is the exponential variate, $N(0,1)$ is the normal random variate with zero mean and variance unity,
%and $R(0,1)$ is the uniform random variate in the interval $[0,1]$.
%In the above equations, $\gamma$ and $\delta$ can be chosen in any combination
%such that the desired $\alpha = \delta/\gamma$.

\section{Fitting the MLP function to the IMF} \label{FitSec}

\begin{figure}
\centering
\includegraphics[width = 0.47\textwidth]{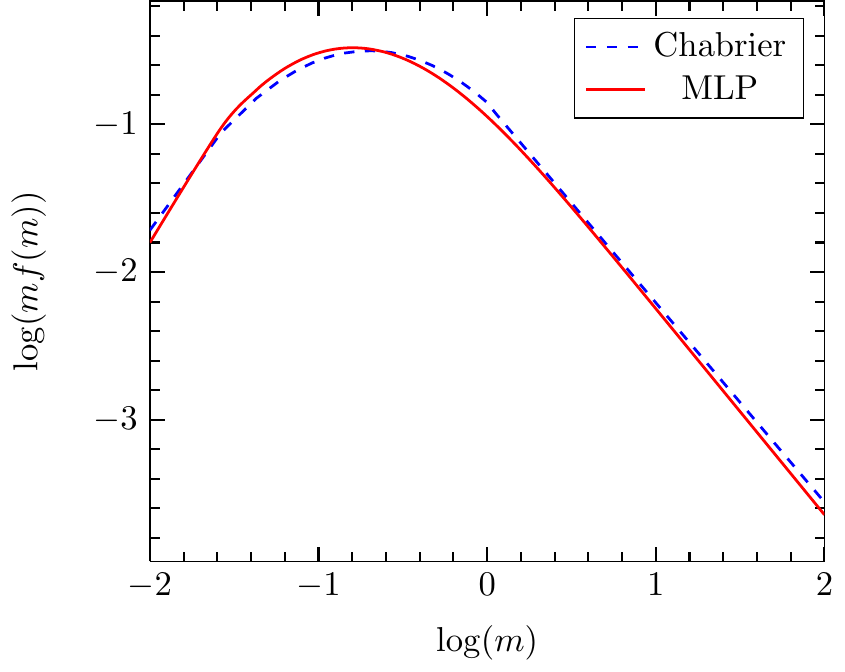}
\caption{\label{MLPFitsFig}Comparison of the Chabrier IMF with its corresponding best fit MLP, where $\mu_0 = -2.404$, $\sigma_0 = 1.044$, and $\alpha = 1.396$. The mass is normalized by $M_{\odot}$. }
\end{figure}

\begin{figure}
\centering
\includegraphics[width = 0.47\textwidth]{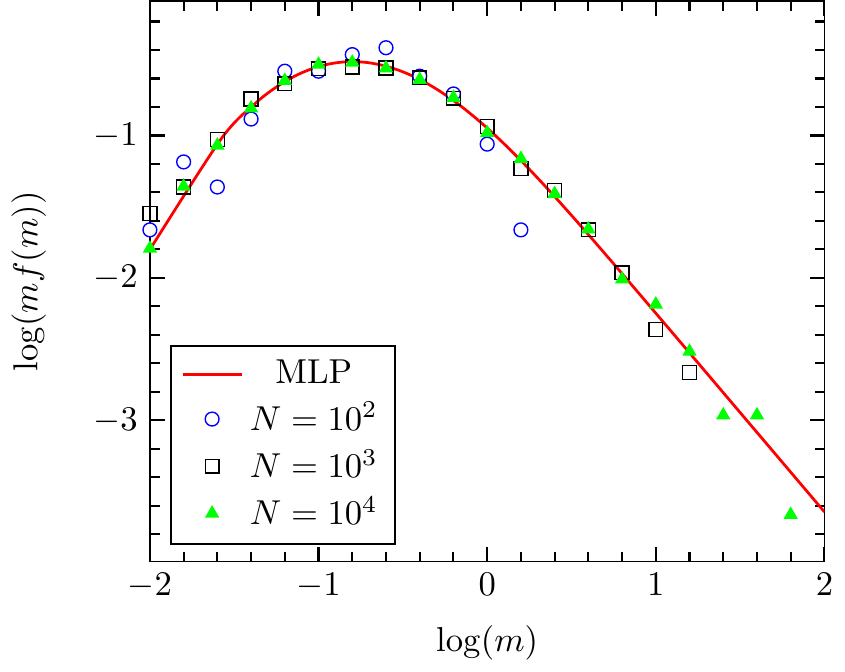}
\caption{\label{MLPThreeSamplesFig}The MLP function wth $\mu_0 = -2.404$, $\sigma_0 = 1.044$, and $\alpha = 1.396$, overlaid with histogram values for random samples drawn from the distribution with size $N=100$ (blue circles), $N=1000$ (black squares), and $N=10000$ (green triangles). All histograms are binned in increments $\Delta \log m = 0.2$, and histogram values are the fractional number in each bin divided by $\Delta \ln m$.} 
\end{figure}

Here, we find a set of parameters for the MLP distribution that fit the updated
IMF of \citet{Chabrier2}. We use 
a normalized form of Equation (1) in \citet{Chabrier2}. It is divided into
logarithmic bins of mass normalized by $M_{\odot}$ with
intervals of 0.05 over the range of $0.06 - 100\,M_{\odot}$ and we use a Levenberg-Marquardt least-squares minimization method to fit the MLP function.
The best fit parameters are $\mu_0 = -2.404$, $\sigma_0 = 1.044$, and
$\alpha = 1.396$. \autoref{MLPFitsFig} shows the best fit MLP function along with the 
Chabrier function, both in normalized form. 
\autoref{MLPThreeSamplesFig} shows three sets of random samples drawn from this best fit MLP function,
using \autoref{mlpvar} and sample sizes of 100, 1000, and 10000, respectively. The samples are
made into histograms with logarithmic bin width of $0.2$, and each histogram is represented by a 
different symbol. We find that samples of well over 100 are needed to adequately populate the power-law tail. For MLP samples of about 100 or less, fitting a lognormal pdf typically yields a better goodness-of-fit statistic than the MLP distribution, given that it also has one less fitting parameter. 

Some sample routines for fitting the MLP function
or drawing random samples from it can be found at \href{http://www.astro.uwo.ca/~basu/mlp.html}{\tt http://www.astro.uwo.ca/$\sim$basu/mlp.html}. 

With a best fit MLP function in hand, it is relatively easy to quantify several important properties of the IMF. \autoref{mean} can be used to calculate the mean mass in the distribution; it is $0.55\,M_{\odot}$. If we were to truncate the distribution at $100\,M_{\odot}$, the mean mass is about 10\% lower at $0.49\,M_{\odot}$. The mode of the distribution $f(m)$ is calculated numerically to be $0.04\,M_{\odot}$. However, the usual practice is to plot data as $dN/d\ln M \equiv m\,f(m)$ rather than $dN/dM \equiv f(m)$, so the usually quoted IMF peak is that of $m\,f(m)$. We find this to be $0.16\,M_{\odot}$ for our best fit parameters. The mode of $m\,f(m)$ is found analogously to that for $f(m)$ described in \autoref{mode}, but with $\alpha$ lowered by one integer value. 

\begin{figure}
\centering
\includegraphics[width = 0.47\textwidth]{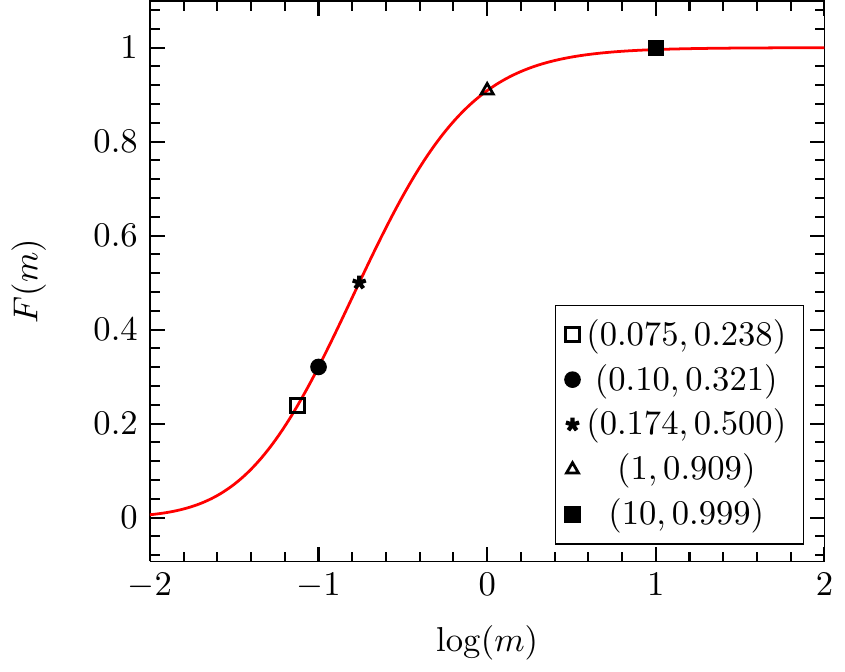}
\caption{\label{cumFig}MLP cumulative distribution with $\mu_0 = -2.404$, $\sigma_0 = 1.044$, and $\alpha = 1.396$. The mass is normalized by $M_{\odot}$. The legend illustrates interesting values of the pairs $(m, F(m))$.}
\end{figure}

\begin{figure}
\centering
\includegraphics[width = 0.47\textwidth]{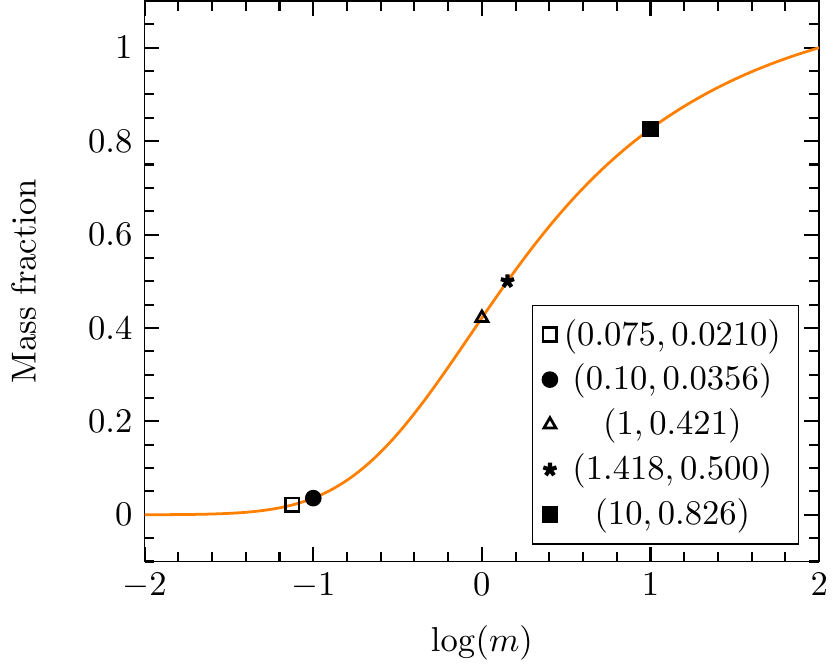}
\caption{\label{massFracFig} Mass fraction for an MLP distribution with $\mu_0 = -2.404$, $\sigma_0 = 1.044$, and $\alpha = 1.396$. The mass is normalized by $M_{\odot}$. The legend illustrates interesting values of pairs of $m$ and the mass fraction.}
\end{figure}

\autoref{cumFig} shows the cumulative MLP distribution for the best fit parameters. The legend panel in \autoref{cumFig} illustrates the specific values for $F(m)$ for several relevant mass values, e.g., one solar mass, as well as the minimum stellar mass, $m = 0.075 M_{\odot}$ \citep{Chabrier3}. Although the data used to generate the IMF quoted by \citet{Chabrier2} does not span the entire hypothetical mass range $m \in [0,\infty]$, one can use the numbers in \autoref{cumFig} to quickly estimate some highlights of the mass function, with interesting pairs of $m,F(m)$ marked by symbols in \autoref{cumFig}: about one quarter of objects are substellar, about one third of objects have masses less than $0.1\,M_{\odot}$, the median mass is $0.17\,M_{\odot}$, just over 90\% of objects have masses less than $1\,M_{\odot}$, and only 9\% of objects have masses in the range $1-10\,M_{\odot}$. 
\autoref{massFracFig} shows the mass fraction $\phi(m) = \int_0^m m'\,f(m')\,dm'$ (with closed form given in \autoref{massFracEq}) for the best fit parameters, normalized by the value at $m = 100 M_{\odot}$. Symbols denote interesting pairs of $m,\phi(m)$, and many provide an interesting contrast to their counterparts in \autoref{cumFig}. For example, only about 2\% of the total mass is in substellar objects, a majority of mass is tied up in objects more massive than $1\,M_{\odot}$, and half of all mass is in objects above mass $1.4\,M_{\odot}$.

The ability to model stellar populations with an IMF that has clearly identifiable number or mass fractions in different mass ranges should be a key aid to galaxy studies \citep[e.g.,][]{Kennicutt}, for example in the determination of star formation rates, chemical evolution, and mass-to-light ratios.


\section{Conclusions}\label{conclusions}

We have derived several important properties of the modified lognormal power-law (MLP) probability distribution function that has been recently introduced in the literature. The three-parameter MLP function has the salutary properties of a main body that resembles a lognormal, including a peak value and a decline toward low values, as well as a power-law tail at high values. This
function can potentially be applied in a variety of fields where empirical distributions may have a power-law tail that coexists with a peak at lower values. The MLP distribution can also help to settle a frequent contentious question: is a data set consistent with a lognormal distribution or does it also show evidence for a power-law tail? Simultaneous fitting of the lognormal and MLP distributions to the same data sets can help to answer this question in many fields.

Comparison of real data sets with the MLP distribution is facilitated by the results presented in this paper. We have derived analytic expressions for the cumulative distribution function, the mean, variance, and higher moments of the distribution. We also derived an approximation for the mode that can serve as an initial guess for nonlinear solvers that can iterate to a more exact solution. The random variate of the MLP distribution has also
been introduced. Together, these results can put the MLP distribution on a more equal footing with many classical distributions (e.g., lognormal, exponential, Pareto, Rayleigh) that are frequently used to fit empirical data. The use of the MLP distribution
to fit an empirical data set has been demonstrated and we show how useful information about the mean, mode, median, and distributions of number or mass fractions can be easily calculated.

\section{Acknowledgements}\label{acknowledgments}
SB acknowledges the hospitality of the Isaac Newton Institute of Mathematical Sciences at Cambridge University during the initial stages of writing of this paper. This work was supported by an NSERC Discovery Grant to SB. 

%\bibliographystyle{mn2e}
%\bibliography{SEPT292014MLP}

\appendix

\section{Error Function}
\subsection{Definition and Basic Properties}
\begin{figure}
	\centering
	\includegraphics[width = 0.47\textwidth]{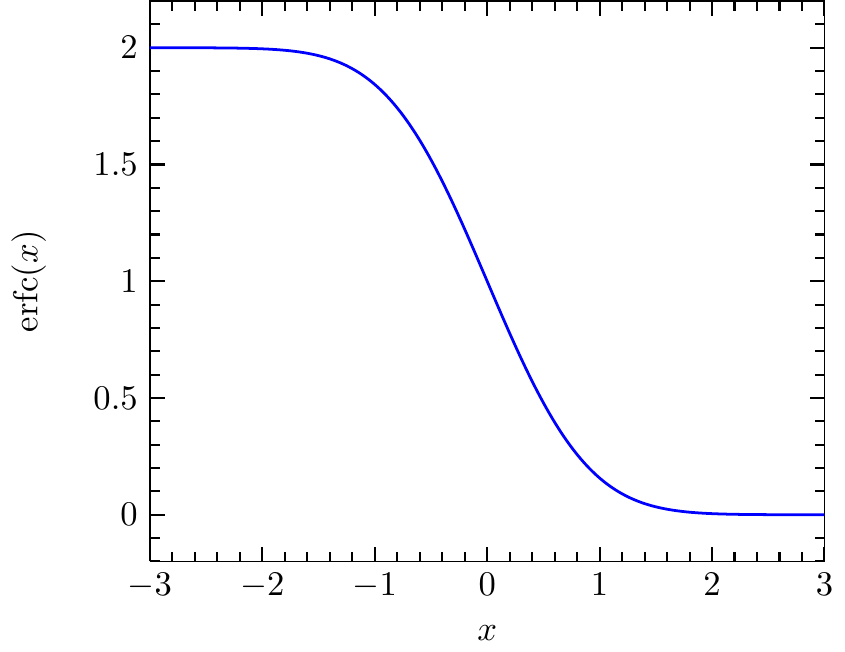}
	\caption{\label{erfcFig} Complementary error function, $\erfc(x)$. }
\end{figure}

The Error function, denoted by $\erf(x)$, and the Complementary Error Function, denoted by $\erfc(x)$, are defined as
\beg \label{erf} \erf(x) = \frac{2}{\sqrt{\pi}}\int_0 ^ x e^{-t^2}dt \ , \e
\beg \label{erfc} \erfc(x) = 1 - \erf(x) = \frac{2}{\sqrt{\pi}}\int_x^\infty e^{-t^2}dt \ . \e
Note from the definition that $\erf(x)$ is an odd function.
Also,
\beg \label{erfLim}\underset{x\rightarrow \infty}\lim \erf(x) = 1 \  \e
\citep{Abramowitz}. The derivative and integral of $\erf(x)$ are:
\beg \label{erfDder} \frac{d}{dx}\erf(x) = \frac{2}{\sqrt{\pi}}e^{-x^2} \ , \e
\beg \label{erfInt} \int \erf(x) dx = x\,\erf(x) + \frac{1}{\sqrt{\pi}}e^{-x^2} + C \ , \e
where $C$ is an integration constant. Finally, the Taylor expansion for $\erf(x)$ is given by 
\beg \label{erfTaylor} \erf(x) = \frac{2}{\sqrt{\pi}}\underset{n=0}{\overset{\infty}{\sum}}\frac{(-1)^n x^{2n +1}}{(2n+1)n!} = \frac{2}{\sqrt{\pi}}\left(x -\frac{x^3}{3} + \frac{x^5}{10} + ... \right) \ .\e
A plot of $\erfc(x)$ is presented in \autoref{erfcFig}. 
\subsection{Related Integrals used in our analysis}
Consider the integral
\beg I_1 =  \int \exp[-(ax^2 + bx + c)]dx \, , \ a > 0 \e

By completing the square and letting $u = \sqrt{a}\left(x+\frac{b}{2a}\right)$ it can be brought to
\begin{align} I_1 = \label{I1}  \frac{1}{2}\sqrt{\frac{\pi}{a}}\exp\left(\frac{b^2 - 4ac}{4a}\right)\erf\left(\sqrt{a}\left[x + \frac{b}{2a}\right]\right) + \tilde{C}\ ,\end{align}
where $\tilde{C}$ is an integration constant. In particular, using the properties of $\erf(x)$ and its relation to $\erfc(x)$, we have
\begin{align} \label{I11} \int _0 ^\infty &\exp[-(ax^2 + bx + c)]dx \notag \\
&= \frac{1}{2}\sqrt{\frac{\pi}{a}}\exp\left(\frac{b^2 - 4ac}{4a}\right)\erfc\left(\frac{b}{2\sqrt{a}}\right) \ .\end{align}

The result in \autoref{I1} can also be used to find integrals of the form
\beg I_2 =  \int x^{-(\alpha + 1) }\exp\left[-\left(B+ C\ln x\right)^2 \right]dx \ , \e
where $C\neq 0$ and $x >0$. By using the substitution $v = B + C\ln x$, we can bring it to the form
\beg I_2 = \frac{1}{C}\int \exp\left[-\left(v^2 +v\frac{\alpha}{C} - \frac{\alpha B}{C}\right)\right]dv \ , \e
which we can evaluate to
\beg \label{I2}  I_2 = \frac{\sqrt{\pi}}{2C}\exp\left(\frac{\alpha^2 + 4\alpha BC}{4C^2}\right)\erf\left(B + C\ln x + \frac{\alpha}{2C}\right) +  \tilde{C} \ . \e
\section{Derivation of the Results in section 5}\label{Derivations}

\subsection{\label{pdf}The MLP as a Probability Density Function}
We can explicitly verify that the MLP density $f(m)$ (\autoref{MLP}) is indeed a valid probability density function for all $\alpha >0$, i.e., it is positive and integrates to 1. Positivity is clear from the definition. To verify the normalization condition, let $A = (\alpha/2)\exp\left(\alpha\mu_0 +\alpha ^2\sigma_0 ^2 / 2\right)$, $B = (1/\sqrt{2})\left[\alpha\sigma_0 + (\mu_0/\sigma_0)\right]$, and $C = -1/(\sqrt{2}\sigma_0)$. We then have
\beg \int _0 ^{\infty}f(m)dm =  A\int_0^\infty m^{-(1+\alpha)} \erfc(B + C\ln m) dm \ . \label{MLPNormInt} \e
Letting $dv = m^{-(\alpha + 1)}dm$ and  $ u = \erfc(B + C \ln m)$, we can write this integral as

\begin{align} &\int _0 ^{\infty}f(m)dm =  \left. -\frac{A}{\alpha}m^{-\alpha}\erfc(B + C \ln m) \right|_0^\infty \notag\\ 
&- \frac{2AC}{\alpha\sqrt{\pi}}\int _0^\infty m^{-(\alpha + 1)}\exp\left[-\left(B+ C\ln m\right)^2 \right] dm \ . \label{MLPNormInt2}\end{align}
Consider the first term on the right hand side in \autoref{MLPNormInt2}. As $m\rightarrow \infty$, $\erfc(B + C \ln m) \rightarrow 2$ (since $C<0$ and $\erf(x)$ is odd), hence for $\alpha>0$ the term vanishes in this limit. As $m\rightarrow 0$, the limit takes on an indeterminate form. Applying L'Hospital's rule we can see that the limit is also zero in this case:
\begin{align} \label{lim}\underset{m\rightarrow 0}{\lim}& \ m^{-\alpha}[\erfc(B + C\ln m)] \notag\\ 
&=  -\frac{2C}{\sqrt{\pi} \alpha}\underset{m\rightarrow 0}{\lim}\frac{e^{-(B + C\ln m)^2}}{m^{\alpha}} \notag \\ &= -\frac{2C}{\sqrt{\pi}\alpha}\underset{x \rightarrow \infty}{\lim}\exp\left[-x^2 - \frac{\alpha}{C}(x-B)\right] = 0 \ ,\end{align}
where $x = B + C\ln m$. We can evaluate the second (integral) term with \autoref{I2}: 

\begin{align} &-\frac{2CA}{\alpha\sqrt{\pi}}\int_0 ^\infty m^{-(\alpha + 1) }\exp\left[-\left(B+ C\ln m\right)^2 \right]dm \notag \\
&= \left.-\frac{A}{\alpha}\exp\left(\frac{\alpha^2 + 4\alpha BC}{4C^2}\right) \times \erf\left(B + C\ln x + \frac{\alpha}{2C}\right) \ \right| _0 ^\infty \notag \\
&= \frac{2A}{\alpha}\exp\left(\frac{\alpha^2 + 4\alpha BC}{4C^2}\right) = 1 \ . \label{momInt}\end{align}

\subsection{Cumulative Distribution}

The cumulative distribution function, %
\beg F(m) = \int_{-\infty}^m f(t)dt \ ,\e 
and for our case $f(m)$ is defined on $[0,\infty)$. To find the closed form we first we apply integration by parts, using also the same definitions for $A$, $B$, and $C$ as in \autoref{pdf}:
\begin{align} F(m) = &\left. -\frac{A}{\alpha}t^{-\alpha}\erfc(B + C \ln t)\right|_0^m \notag \\&- \frac{2AC}{\alpha\sqrt{\pi}}\int_0 ^m t^{-(\alpha + 1)}\exp\left[-\left(B+ C\ln t\right)^2 \right] dt \ . \end{align} 
We again use the identity in \autoref{I2} to evaluate the integral term and obtain
\begin{align} F(m) &= -\frac{A}{\alpha}m^{-\alpha}\erfc(B + C \ln m) \notag \\& \hspace{-10pt} + \frac{A}{\alpha}\exp\left(\frac{\alpha^2 + 4\alpha BC}{4C^2}\right)\erfc\left(B + C\ln m + \frac{\alpha}{2C}\right)\, , \label{preCumulative}
\end{align}
which, upon returning to the original parameters, becomes
\begin{align} F(m) &= \frac{1}{2}\erfc\left(-\frac{\ln(m)-\mu_0}{\sqrt{2}\sigma_0}\right) \notag\\
&\hspace{-20pt}-\frac{1}{2}\exp\left(\alpha\mu_0+\frac{\alpha^2\sigma_0^2}{2}\right)m^{-\alpha}\erfc\left(\frac{\alpha\sigma_0}{\sqrt{2}}-\frac{\ln(m)-\mu_0}{\sqrt{2}\sigma_0}\right) .
\end{align}

\subsection{Raw Moments}

Next we derive a closed form for arbitrary raw moments of the distribution, as well as an expression for its variance. Let $M$ be an MLP random variable with probability density function $f(m)$. The $k$th raw moment of $M$, defined as the expectation value of $M^k$, is given by
\begin{align} \notag E[M^k] &= \int_0 ^\infty m^k f(m)dm \\ \label{moment} &= A\int_0^\infty m^{k-(\alpha + 1)}\erfc(B + C\ln m)dm \ , \end{align}
with $A, B$, and $C$ as defined in \autoref{pdf}. Before we arrive at a closed form expression for the moments we consider the convergence of the integral in \autoref{moment}. This integral diverges for $k \geq \alpha$. To see this, note that if $k \geq \alpha$ then $p = \alpha + 1 - k \leq 1.$ Now write \autoref{moment} as

\begin{align}\label{moment2} 
\int_0 ^\infty \frac{\erfc(B + C\ln m)}{m^p}dm &= \int_0 ^a \frac{\erfc(B + C\ln m)}{m^p}dm \notag\\ &+ \int_a ^\infty \frac{\erfc(B + C\ln m)}{m^p}dm \ , \end{align}    
where $a \in (0,\infty)$ such that $\erfc(B + C\ln m) \geq 1, \ \forall m \geq a$. The existence of such $a$ is ensured by the fact that for $C<0$,  $\erfc(B + C\ln m)$ is a continuous strictly increasing positive function having an upper limit of 2. Then 

\beg \int_a^\infty \frac{1}{m^p}\erfc(B + C\ln m)dm \geq \int_a^\infty \frac{1}{m^p} dm \ ,\e
where the integral on the right hand side diverges for $p \leq 1$. Thus, the integral in \autoref{moment} diverges for $k\geq \alpha$. Conversely, the moments are finite for $k< \alpha $. For any $a \geq 0$, 

\beg \int_a^\infty \frac{1}{m^p}\erfc(B + C\ln m)dm \leq 2\int_a^\infty \frac{1}{m^p} dm \ , \e
which converges for $p = \alpha - k + 1 > 1$. Together with \autoref{moment2} and \autoref{lim} this proves the existence of the moments.
Suppose that $\alpha > k$.  Then by simply letting $\alpha \rightarrow \alpha - k > 0$ in the integrand of \autoref{MLPNormInt}, we can see from \autoref{lim} and 	\autoref{momInt} that the $k$th moment is given by 
\begin{align} E[M^k] &= \frac{2A}{\alpha - k}\exp\left(\frac{(\alpha - k)^2 + 4(\alpha - k)BC}{4C^2}\right) \notag \\ &=\frac{\alpha}{\alpha - k} \exp\left(\frac{\sigma_0^2k^2}{2} + \mu_0 k\right) \ . \end{align}

\subsection{Mass Fraction}

We can obtain the expression for the mass fraction (up to a mass $m$), 
\begs \phi(m) = \int_0^{m} m'f(m') dm' = A\int_0^m (m')^{-\alpha} \erfc(B + C\ln m') dm' , \es
using the same approaches employed in the computation of the cumulative distribution and the raw moments in the previous sections. Since the mean exists only if $\alpha > 1$, it is also natural to consider the closed-form expression for the mass fraction for $\alpha > 1$. This can be achieved by replacing all explicit appearances of $\alpha$ with $\alpha-1$ in \autoref{preCumulative}, while maintaining the dependencies in the parameters $A(\alpha)$, $B(\alpha)$ and $C(\alpha)$ in the original form, as we did for the calculation of the raw moments. Thus,
\begin{align} \phi(m) = &-\frac{A}{\alpha-1}m^{-(\alpha-1)}\erfc(B + C \ln m) \notag \\& + \left\{ \frac{A}{\alpha-1}\exp\left(\frac{(\alpha-1)^2 + 4(\alpha-1) BC}{4C^2}\right)\right. \notag\\
&\hspace{20pt}\times \left. \erfc\left(B + C\ln m + \frac{\alpha-1}{2C}\right) \right\}, \alpha > 1, 
\end{align}

which, upon returning to the original parameters, becomes
\begin{align} \phi(m) &= \frac{1}{2}\frac{\alpha}{\alpha-1}\exp\left(\frac{\sigma_0^2}{2} + \mu_0\right)\erfc\left(\frac{\sigma_0}{\sqrt{2}} - \frac{\ln m - \mu_0}{\sqrt{2}\sigma_0}\right)\notag\\
& - \left\{\frac{1}{2}\frac{\alpha}{\alpha-1}\exp\left(\frac{\alpha^2\sigma_0^2}{2} + \alpha\mu_0\right)m^{-(\alpha-1)}\right. \notag\\
 & \left. \hspace{20pt}\times \erfc\left(\frac{\alpha\sigma_0}{\sqrt{2}} - \frac{\ln m - \mu_0}{\sqrt{2}\sigma_0}\right)\right\}, \alpha > 1. \label{massFracEq}
\end{align}
Note that, as expected, $\phi(m)$ becomes the expression for the expectation value as $m \rightarrow \infty$ (for $\alpha > 1$).

\end{document}